# Spin- and orbital-angular-momentum nonlinear optical selectivity of single-mode nanolasers

Chenglin He[1], Zilan Tang[1], Liang Liu[1], Stefan A. Maier[2,3,4],

Xiaoxia Wang[1,*], Haoran Ren[2,*], Anlian Pan[1,*]

[1]Hunan Institute of Optoelectronic Integration and Key Laboratory for MicroNano Physics and Technology of Hunan Province, State Key Laboratory of Chemo/Biosensing and Chemometrics, College of Materials Science and Engineering, Hunan University, Changsha 410082, P. R. China.

[2]School of Physics and Astronomy, Faculty of Science, Monash University, Melbourne, Victoria 3800, Australia.

[3]Department of Physics, Imperial College London, London, SW7 2AZ, UK.

[4]Chair in Hybrid Nanosystems, Nanoinstitute Munich, Faculty of Physics, Ludwig Maximilian University of Munich, Munich 80539, Germany.

*Emails: wangxiaoxia@hnu.edu.cn; Haoran.Ren@monash.edu; Anlian.Pan@hnu.edu.cn.

**Abstract**

Selective control of light is essential for optical science and technology with numerous applications. Nanophotonic waveguides and integrated couplers have been developed to achieve selective coupling and spatial control of an optical beam according to its multiple degrees of freedom. However, previous coupling devices remain passive with an inherently linear response to the power of incident light limiting their maximal optical selectivity. Here, we demonstrate nonlinear optical selectivity through selective excitation of individual single-mode nanolasers based on the spin and orbital angular momentum of light. Our designed nanolaser circuits consist of plasmonic metasurfaces and individual perovskite nanowires, enabling subwavelength focusing of angular-momentum-distinctive plasmonic fields and further selective excitation of single transverse laser modes in nanowires. The optically selected nanolaser with nonlinear increase of light emission greatly enhances the baseline optical selectivity offered by the metasurface from about 0.4 up to near unity. Our demonstrated nonlinear optical selectivity may find important applications in all-optical logic gates and nanowire networks, ultrafast optical switches, nanophotonic detectors, and on-chip optical and quantum information processing.

**Introduction**

The ability to select and control light through its multiple degrees of freedom is of great importance for widespread photonic applications towards higher dimensions. Compact nanophotonic waveguides and integrated couplers have been developed to achieve wavelength (*1, 2*) and polarization (*3, 4*) selectivity. Moreover, selective routing of spin angular momentum (SAM) carried by circularly polarized photons (*5-7*) and valley emissions (*8-10*) have been realized from nanophotonic waveguides via spin-orbit coupling. Twisted light carrying an unbounded set of orbital angular momentum (OAM) modes holds great promise for scaling up photonic capacity in optical communications (*11, 12*), holographic displays (*13*), data storage (*14, 15*), and quantum cryptography (*16*). Recently, OAM selectivity has been realized from plasmonic couplers (*17-19*) and silicon waveguides (*20*). However, previous nanophotonic coupling devices, developed for the selective control of light, remain passive with an inherently linear power dependence on incident light, limiting their maximal optical selectivity.

Nanolasers — a new class of amplifiers and integrated light sources in nanophotonic circuits —are of paramount importance for on-chip photonics and all-optical processing (*21*). Nanolasers exploit resonant enhancement of non-radiative cavity modes for stimulated emission, the nonlinear power dependence of which can potentially enhance optical selectivity by leveraging optical contrast below and above the lasing threshold (*22*). High-gain nanolasers made of different gain mediums of II-VI (*23, 24*) and III-V dielectric compounds (*25, 26*) have been designed in conjunction with different dielectric resonators, including photonic crystals (*27*), Fabry-Pérot cavity (*24*), and whispering-gallery-mode cavities (*28*). Although all-dielectric nanolasers feature low loss and high-quality-factors, hybrid and plasmonic nanolasers amplify surface plasmons and embrace the capability of nanoscopic mode localization beyond the optical diffraction limit (*29*). Nevertheless, it is still elusive to realize selective excitation of nanolasers based on optical degrees of freedom.

Here we demonstrate a new concept of nonlinear optical selectivity using hybrid plasmonic nanolaser circuits exhibiting near-unity optical selectivity by the SAM and OAM of incident light (Fig. 1). Our nonlinear-optical-selectivity nanolaser circuits (NNCs) consist of judiciously designed plasmonic metasurfaces on a thin silver plate and pairs of well-positioned perovskite ($CsPbBr_3$) nanowires (Fig. 1A). The plasmonic metasurfaces are designed to convert and focus incident SAM ($s$=-1 and +1) and OAM ($l=l_1$ and $l_2$) into spatially distinctive plasmonic fields with subwavelength confinement. The tightly confined plasmonic fields are further coupled to the end faces of individual nanowires, capable of selective excitation of single laser modes in the high-optical-gain nanowires (Fig. 1B). Unlike far-field optical excitation of multiple longitudinal photonic modes in perovskite nanowire lasers on a dielectric substrate (*30*), the evanescent plasmonic fields can excite transverse Fabry–Pérot (FP) cavity modes of the nanowire for single-mode lasing (inset of Fig. 1B). Strikingly, owing to the nonlinear power dependence in nanowires by transitioning from spontaneous emission to stimulated emission, the base optical selectivity offered by the plasmonic metasurfaces of NNCs is significantly enhanced from only about 0.4 up to near unity (Fig. 1C). Therefore, our hybrid NNCs unveil a new nanophotonic platform to achieve the maximum optical selectivity.

**Results and discussion**

The principle of subwavelength excitation of single-mode nanowire lasers is demonstrated in Fig. 2. Without loss of generality, we design a semi-circular plasmonic lens engraved on a 200 nm-thick silver plate to excite a plasmonic focal spot, which is further coupled to a well-positioned perovskite ($CsPbBr_3$) nanowire for single-mode lasing (Fig. 2A). The periodic semi-circular grooves in the plasmonic lens were optimized to have a pitch of 450 nm, a width of 150 nm, and a depth of 200 nm, offsetting the momentum mismatch between the surface plasmon polaritons (SPPs) on the silver/air interface and the incident light of wavelength 470nm (Fig. S1). The plasmonic lens leads to a subwavelength focus with the full-width-at-half-maximum (FWHM along the x axis in Fig. 2A) below 150 nm, paving the way for subwavelength excitation of the nanowire (Fig. S2). We show that the coupling efficiency is highest by placing the end face of the nanowire near the plasmonic focus (Figs. 2B and S3) with an ultrasmall interaction volume of around $\lambda_0^3/1200$ and a large field enhancement, which can effectively excite the transverse F-P cavity mode of the nanowire (Fig. 2C).

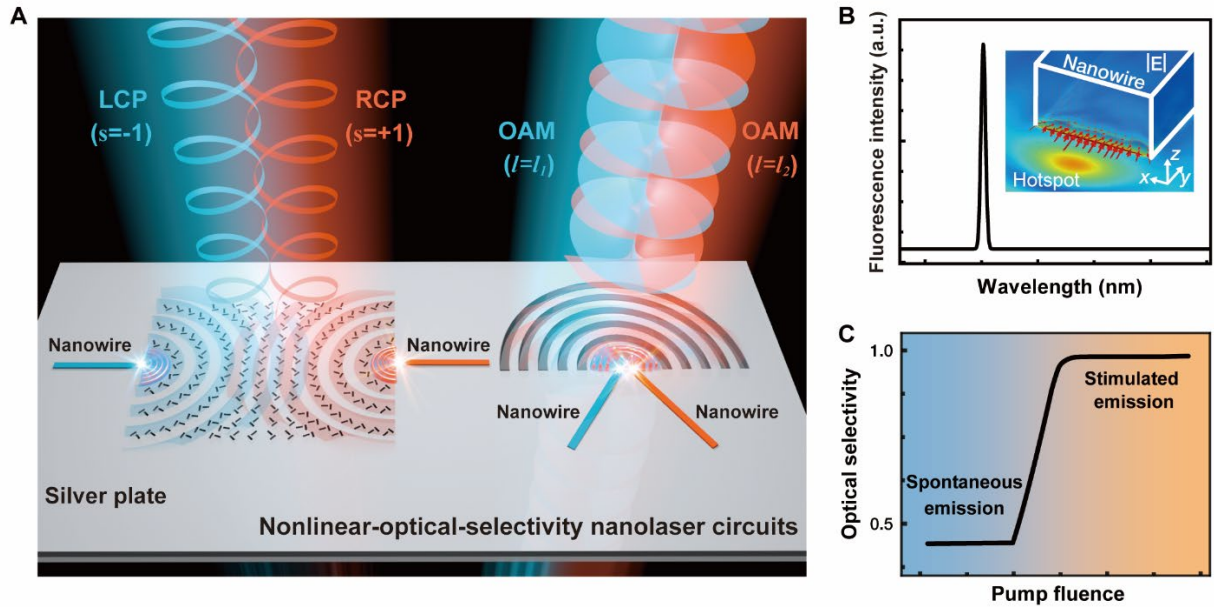

**Fig. 1. Principle of nonlinear optical selectivity achieved from hybrid nanolaser circuits.** (**A**) Schematic representation of the designed nonlinear-optical-selectivity nanolaser circuits (NNCs). The NNCs consist of plasmonic metasurfaces and well-positioned individual perovskite nanowires, which are used for converting the incident SAM with s=±1 (carried by the left- and right-handed circular polarization (LCP and RCP) and OAM with different topological charges ($l_1$ and $l_2$) into spatially distinctive plasmonic fields which further excite stimulated laser emission in respective nanowires. (**B**) Single-mode lasing spectrum from a perovskite nanowire, which originates from subwavelength excitation of transverse Fabry–Pérot cavity mode in the nanowire (inset). (**C**) Boosting optical selectivity through the nonlinear lasing process in the NNCs, wherein the power increase of a pump laser allows the transition from spontaneous emission, to amplified spontaneous emission, and to stimulated laser emission.

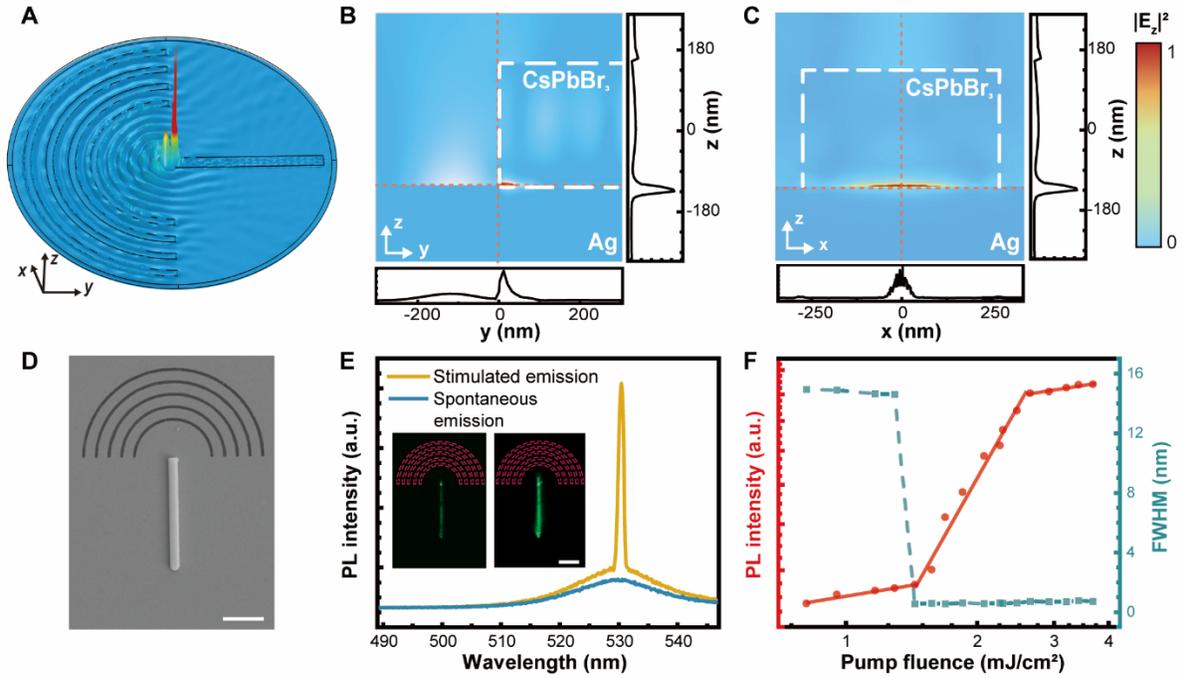

**Fig. 2. Principle of subwavelength excitation of single-mode nanolasers.** (**A**) Electric field distribution (3D profile) of a strongly confined plasmonic focal field located at one of the end faces of a dielectric nanowire. (**B** and **C**) Electric field distributions (2D profiles) of the plasmonic focus in the nanowire-coupled area in the *y-z* (B) and *x-z* (C) planes, which are extracted from the simulation in (A). (**D**) Scanning electron microscopy (SEM) image of one of fabricated hybrid plasmonic nanolaser devices. The scale bar is 2 μm. (**E**) Measured photoluminescence (PL) spectra from the nanowire sample in (D), with the incident laser power below (blue) and above (yellow) the nanowire lasing threshold. The insets present the corresponding dark-field microscope images. The scale bar is 3 μm. (**F**) Measured PL intensity (red) and the full-width-at-half-maximum (FWHM, green) of the PL spectra as a function of the pump fluence, yielding a lasing threshold of 1.4 mJ/cm$^2$.

To experimentally confirm the subwavelength excitation of single-mode nanolasers, we fabricate the designed plasmonic lens via focused ion beam lithography on a silver film deposited on an ultraflat quartz substrate. Thereafter, CsPbBr$_3$ nanowires synthesized by chemical vapor deposition method are precisely transferred to the desired target positions via a micro-manipulation technique (Fig. S4). The scanning electron microscope (SEM) image of one of fabricated devices is given in Fig. 2D, wherein we choose a nanowire with a lateral size of ~500 nm to verify the single-mode lasing. A femtosecond laser of pulse repetition rate of 1kHz and wavelength of 470 nm is prepared to have a circular polarization state (Fig. S5). To avoid direct excitation of nanowires from the incident light, the nanowire sample is illuminated from the opposite (substrate) side of the silver film by a low numerical aperture lens (NA=0.2, 10x), ensuring that nanowire excitation is only possible from evanescent plasmonic fields. To achieve high collection efficiency, the nanowire photoluminescence (PL) emission is collected in transmission by a high-NA lens (NA=0.95, 100x). Initially, a broad PL spectrum originating from spontaneous emission in the nanowires is detected when the incident power is kept low (Fig. 2E). By increasing the pump fluence, the nanowire transits from spontaneous emission, to amplified spontaneous emission, and finally to stimulated laser emission, which is accompanied with a sharp decrease in the linewidth of the PL spectra centered at 530 nm (Fig. 2E, 2F and S6). We compare the free-spectral ranges of the longitudinal and transverse F-P cavity modes in the nanowire (Fig. S7), suggesting that our observed single-mode lasing originates from the transverse F-P cavity modes. In addition, we show that the transverse width of nanowires can affect mode coupling efficiency and lead to different lasing thresholds (Fig. S8). The subwavelength excitation of single-mode nanowire lasers lays the foundation for boosting optical selectivity through the rapid and nonlinear increase of PL emission. Next, we design our first NNCs sample based on a chirality-selective plasmonic metasurface consisting of an array of mutually perpendicular nanoslit pairs to achieve the SAM-dependent directional coupling and focusing of SPPs waves (Fig. 3). Since the SPPs waves launched by the individual nanoslits are anisotropic (only the electric field component perpendicular to the nanoslit can excite SPPs), the initial phase of their launched SPPs waves toward the left and right directions is different. This relative phase can be controlled by the spatial arrangement of nanoslits and incident polarization, which determines the interference and directional coupling of the surface waves (Fig. 3A). For the sake of simplicity, we consider identical nanoslits with orthogonal orientations $|\theta_1 - \theta_2| = 90°$ and fix their spacing distance $d$ as one quarter of the SPPs wavelength ($d = \lambda_{SPPs}/4$). This leads to unidirectional launching of the SPPs waves propagating to the left and right sides of the nanoslit pairs based on the incident SAM of $s=-1$ and $s=+1$, respectively (Fig. 3A). To obtain a higher degree of the SAM selectivity, the nanoslit dimensions are further optimized to have a length $L$ of 250 nm, a width $W$ of 80 nm, and a spacing $d$ of 110 nm (Fig. S9). Spatially arranging an array of such nanoslit pairs into a curved shape (Fig. 3B) leads to the design of a chirality-selective plasmonic metasurface, allowing the SPPs fields excited from LCP and RCP incidence being selectively focused to the left and right sides of the metasurface, respectively (Figs. 3C and 3D).

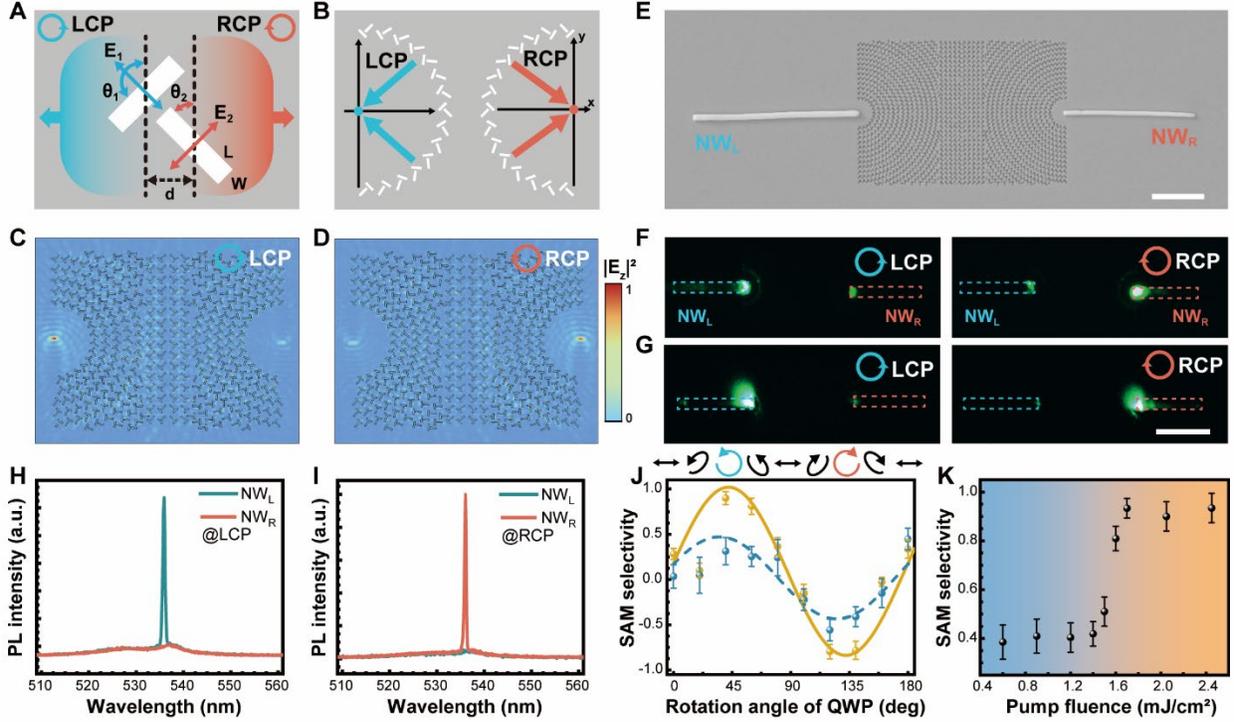

**Fig. 3. Experimental demonstration of nonlinear SAM selectivity in a chirality-selective plasmonic metasurface.** (**A**) Schematic diagram of a nanoslit pair with labelled size parameters. (**B**) Schematic design of a plasmonic metasurface capable of selective focusing of surface waves launched by the incident LCP and RCP beams to the left and right sides, respectively. (**C** and **D**) Top-view electric field distributions of the chirality-selective plasmonic metasurface under the illumination of LCP (C) and RCP (D) incident beams, leading to plasmonic foci on the left and right sides. (**E**) SEM image of the fabricated chirality-selective plasmonic metasurface, with two individual nanowires ($NW_L$ and $NW_R$) transferred and precisely positioned to the plasmonic focal areas in the opposite side. The scale bar is 5 μm. (**F** and **G**) Dark-field microscope images of the chirality-selective plasmonic metasurface under the LCP and RCP incidence with pump fluence below (F) and above (G) the lasing threshold of nanowires. The scale bar is 10 μm. (**H** and **I**) PL emission spectra from the nanowires of $NW_L$ (green) and $NW_R$ (red) under the LCP (H) and RCP (I) light incidence, respectively. (**J**) Circular polarization dependence of the SAM selectivity based on the spontaneous (blue color: below lasing threshold) and stimulated (yellow color: above lasing threshold) emission regimes. The error bars denote the standard deviation of multiple measurements. (**K**) Nonlinear enhancement of the SAM selectivity through pumping the optically selected nanowire above its lasing threshold. The error bars denote the standard deviation of multiple measurements.

To experimentally demonstrate the SAM selectivity, we fabricate the designed chirality-selective plasmonic metasurface with a pitch of 450 nm (Fig. 3E). Two individual nanowires (NW$_L$ and NW$_R$) are transferred and precisely positioned to the plasmonic focal areas in opposite sides. Illuminating the LCP and RCP incident beams on the metasurface, the nanowire PL emission exhibits a baseline SAM selectivity of ~0.47 (Fig. 3F). We define the optical selectivity as $(I_L-I_R)/(I_L+I_R)$, where $I_L$ and $I_R$ represent the integrated PL intensity detected from the left and right nanowires, respectively. Strikingly, we show that the baseline SAM selectivity offered by the metasurface design can be significantly enhanced when the pump fluence is above the lasing threshold in the optically selected nanowires, leading to a near-unity SAM selectivity of ~1 (Fig. 3G). This is mainly due to the rapid and nonlinear increase of light emission in the optically selected nanowires above the lasing threshold, while the counterpart nanowires remain below the lasing threshold with orders of magnitude weaker PL emission.

The nanowire emission spectra under the illumination of LCP and RCP polarizations are presented in Fig. 3H and 3I. All the spectra are measured with the same excitation fluence of ~1.3 $P_{th}$ ($P_{th}$ is the lasing threshold of the nanowire). As expected, LCP and RCP polarizations can selectively excite the single-mode lasing either from $NW_L$ or $NW_R$, showing a sharp spectral response centered at 532 nm. Meanwhile, the other nanowire not optically selected by the circular polarization exhibits a broad and weak spontaneous emission spectrum. Furthermore, we investigate the circular polarization dependence of the SAM selectivity under both the spontaneous and stimulated emission conditions, which is characterized by using the excitation fluences of 0.7 $P_{th}$ and 1.3 $P_{th}$, respectively (Fig. S10). Specifically, rotating the fast axis of a quarter wave plate (QWP) leads to a large dynamic modulation of the degree of circular polarization of incident light from $s$=-1 to +1. We characterize the SAM selectivity of the chirality-selective plasmonic metasurface based on different QWP angles. The resultant optical responses based on the spontaneous and stimulated emission conditions can be fitted with sine functions of ~0.47×sin(2φ) and ~1×sin(2φ), respectively, suggesting that the metasurface working in the stimulated emission regime exhibits a larger modulation range of the SAM selectivity based on the incident polarization. Therefore, we can accomplish an optical-selectivity boost by simply increasing the pump fluence, with which the optically selected nanowire transits from spontaneous emission, to amplified spontaneous emission, and to stimulated laser emission (Fig. 3K).

The principle of nonlinear optical selectivity by single-mode nanolasers can be extended to the OAM degree of freedom of light carried by a helical wavefront (Fig. 4). For this purpose, we design our second NNCs sample based on an OAM-selective nanowire circuit, consisting of semi-circular nanogrooves used for spatially shifting the OAM-dependent plasmonic focal fields on the diameter line of the semi-circles, as well as a pair of perovskite nanowires (Fig. 4A). The distance between the SPPs focal fields excited from adjacent OAM modes is on the order of tens of nanometers (Fig. S11). We transfer a pair of perovskite nanowires ($NW_L$ and $NW_R$) with a splitting angle of 60° to the focal areas of the SPPs fields. The close end faces of the nanowires are displaced by 1.2 μm to be able to select distinctive plasmonic fields excited from different OAM orders (Fig. S12). We investigate the OAM selectivity based on OAM orders ranging from -8 to +8 and measure their respective PL emission spectra at the same excitation fluence of ~1.3 $P_{th}$ (Figs. S13A and S13B). We find strong OAM selectivity by OAM orders of $l$=-4 and +3 for their selective excitation of single-mode lasing in $NW_L$ and $NW_R$, respectively (Figs. S13C and S13D), which is obvious in the dark-field microscope images (Figs. 4B and 4C).

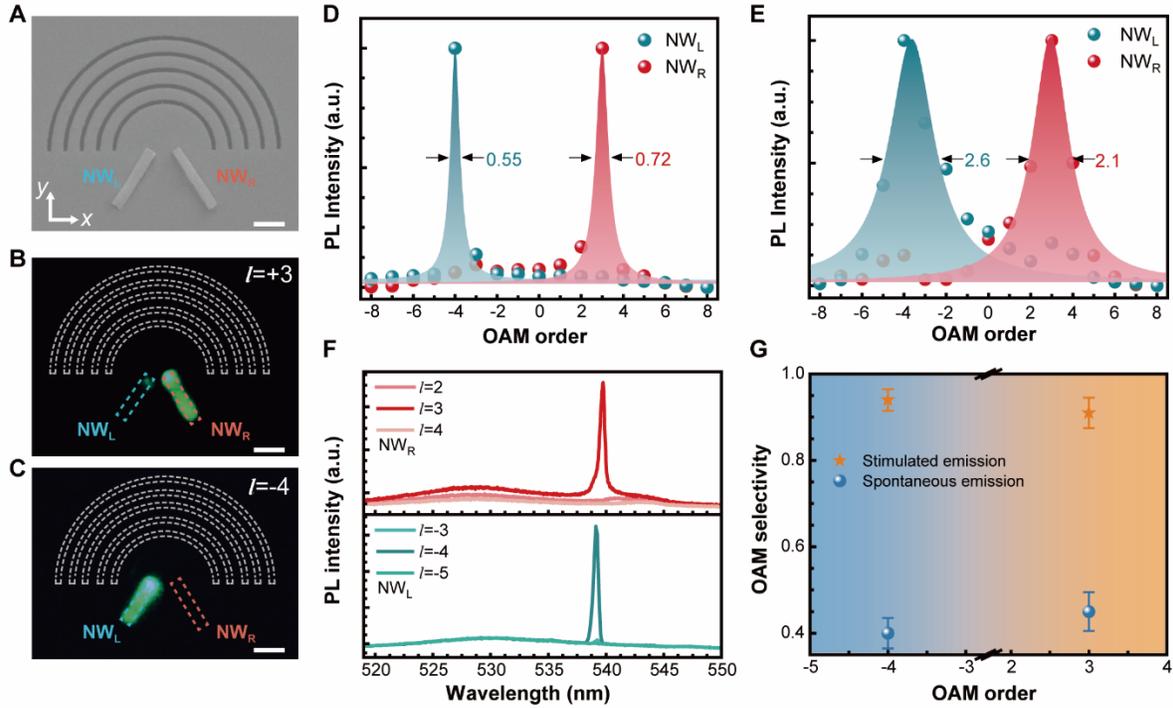

**Fig. 4. Experimental demonstration of nonlinear optical selectivity by an OAM-selective nanowire circuit.** (**A**) SEM image of the fabricated OAM-selective nanowire circuit composed of periodic semi-circular nanogrooves and two individual nanowires ($NW_L$ and $NW_R$). The scale bar is 1 μm. (**B** and **C**) Dark-field optical images of the OAM-selective nanowire circuit under the illumination of the OAM orders of $l$ = -4 (B) and +3 (C), respectively, with the pump fluence above their lasing thresholds. The scale bar is 1 μm. (**D** and **E**) OAM-dependent PL intensity detected from $NW_L$ and $NW_R$ and their Lorentz-fitting curves based on the stimulated (D) and spontaneous (E) emission regimes, respectively. (**F**) PL emission spectra detected from $NW_L$ and $NW_R$ based on adjacent OAM orders. (**G**) Nonlinear enhancement of the OAM selectivity through pumping the optically selected nanowire above its lasing threshold. The error bars denote the standard deviation of multiple measurements.

Impressively, our OAM-selective nanolaser circuit can significantly reduce the modal crosstalk between adjacent OAM modes that exhibit weak PL intensities at the noise level. We characterize the full-width-at-half-maximum (FWHM) of the OAM-dependent PL intensity, giving rise to the results of ~0.55 and ~0.72 for the $NW_L$ and $NW_R$, respectively, both of which are less than one integer number (Fig. 4D). In contrast, when the circuits operate in the spontaneous emission below the lasing threshold, our measured FWHM values are ~2.6 and ~2.1 for $NW_L$ and $NW_R$, respectively (Fig. 4E). To show this reduced crosstalk more clearly, we present the PL emission spectra from $NW_L$ and $NW_R$ based on adjacent OAM orders (Fig. 4F). Since the OAM modal crosstalk is more pronounced for adjacent OAM orders, we define the OAM selectivity as $(I_{l_0} - \frac{\sum I_{l_0 \pm 1}}{2})/(I_{l_0} + \frac{\sum I_{l_0 \pm 1}}{2})$, where $I_{l_0}$ and $\sum I_{l_0 \pm 1}$ represent the integrated PL intensities measured from the desired OAM order and the adjacent OAM orders, respectively. We show that the nonlinear light emission in nanowire lasers can greatly boost the baseline OAM selectivity of ~0.4 to ~0.94 (Fig. 4G). We expect that the OAM selectivity can be further improved up to near unity by employing alternative metasurface design and nanowires with narrower width.

**Conclusion**

In summary, we have demonstrated a metasurface nanophotonic platform to boost optical selectivity, which is achieved from selective excitation of on-chip nanolasers based on the SAM and OAM degrees of freedom of light. We have designed multiple nanolaser circuits consisting of plasmonic metasurfaces and individual perovskite nanowires, capable of selective coupling of the SAM- and OAM-dependent plasmonic focal fields into nanowires for exciting single-mode laser emission. Owing to the nonlinear increase of light emission in nanowire lasers, it opens an unprecedented opportunity to greatly enhance the baseline optical selectivity offered by the metasurface from about 0.4 up to near unity. As such, our NNCs unveil a new nanophotonic platform to achieve the maximum optical selectivity. We believe our demonstrated nonlinear optical selectivity could find important applications in a broad spectrum of photonic applications, such as but not limited to all-optical logic gates, nanowire networks, ultrafast optical switches, nanophotonic detectors, and on-chip optical and quantum information processing.

**Acknowledgements**

**Funding:**

Australian Research Council DECRA Fellowship DE220101085(H. R.)

Australian Research Council Discovery Project DP220102152 (H.R., S.A.M.)

National Natural Science Foundation of China 52221001(A. P.)

National Natural Science Foundation of China 62090035(A. P.)

National Natural Science Foundation of China U19A2090 (A. P.)

National Natural Science Foundation of China 62175061(X.W.)

Key Program of the Hunan Provincial Science and Technology Department 2019XK2001(A. P.)

Key Program of the Hunan Provincial Science and Technology Department 2020XK2001 (A. P.)

Natural Science Foundation of Hunan Province 2022JJ30167(X.W.)

S.A.M. additionally acknowledges the Lee-Lucas Chair in Physics.

**Author contributions**

Conceptualization: X.W., H. R.

Methodology: X.W., H. R., C.H., S. A. M., A. P.

Investigation: C.H., Z.T., L.L., X.W., H. R.,

Visualization: C.H., X.W., H. R.

Funding acquisition: X.W., H. R., S. A. M., A. P.

Project administration: X.W., H. R., A. P.

Supervision: H. R., A. P.

Writing–original draft: C.H., X.W. H. R.

**Competing interests**

All authors declare that they have no competing interests.

**Data Availability**

All data needed to evaluate the conclusions of the paper are available upon reasonable requests.

**Supplementary Materials**

Materials and Methods

Supplementary Text

Figs. S1 to S13

Tables S1

References (1–6)